\documentclass[useAMS,usenatbib,referee]{mn2e}
\usepackage{graphicx}
\usepackage{amsmath}
\title[Probing Mass Segregation in NGC 6397]{Probing Mass Segregation in the Globular Cluster NGC\,6397}

\author[E. Martinazzi, A. Pieres, S. O. Kepler, J. E. S.~Costa, C.~Bonatto, E. Bica]{E.~Martinazzi\thanks{E-mail: elizandra.martinazzi@ufrgs.br}, A.~Pieres, S.~O.~Kepler, J.~E.~S.~Costa, C.~Bonatto, E.~Bica \\
Instituto de F\'{\i}sica, Universidade Federal do Rio Grande do Sul, 91501-900 Porto Alegre, RS, Brazil}

\begin{document}

\date{Accepted 2014 May 23. Received 2014 May 6; in original form 2014 January 13}

\pagerange{\pageref{firstpage}--\pageref{lastpage}} \pubyear{MNRAS 442, 3105–3111 (2014)}

\maketitle

\label{firstpage}

\begin{abstract}

In this study, we present a detailed study of mass segregation in the globular cluster NGC\,6397. First, we carry out a photometric analysis of projected ESO-VLT data (between 1 and 10\,arcmin~from the cluster centre), presenting the luminosity function corrected by completeness. The luminosity function shows a higher density of bright stars near the central region of the data, with respect to the outer region. We calculate a deprojected model (covering the whole cluster) estimating a total number of stars of 193\,000$\pm$19\,000. The shapes of the surface brightness and density-number profiles versus the radial coordinate \emph{r} (instead of the projected coordinate \emph{R}) lead to a decreasing luminosity for an average star, and thus of mass, up to 1\,arcmin, quantifying the mass segregation.
The deprojected model does not show evidence of mass segregation outside this region. 

\end{abstract}


\section{Introduction}

Globular clusters (GCs) are among the oldest structures in the Milky Way. Their fundamental parameters (e.g. age, metallicity, mass, spatial distribution) constitute  a valuable source of basic information on the early stages of Galactic formation \citep[e.g.][]{Gratton2003}. In addition, they can be used as unique test-beds of star formation and stellar evolution models \citep[e.g.][]{2012ApJ...761...51H}.

One of the most interesting aspects of GCs is their dynamical evolution. At high stellar densities, binary encounters may be frequent and, over time, tend to establish a state of equal energy among the member stars. On average, more massive stars transfer orbital energy to the less-massive ones in the encounters. In this context, heavier stars tend to clump in the cluster centre while the lighter ones orbit the outskirts, with some fraction even escaping the system and populating the Milky Way stellar field \citep[e.g.][]{2008ApJ...679.1272M,2012NewA...17..411V}. This dynamical process is called \emph{mass segregation}. Eventually, the accumulation of massive stars in the cluster centre may collapse, thus leading to the \emph{core collapse} phase \citep[e.g.][]{DjorgovskiKing1986,Trager1995}. 

Observationally, mass segregation can be characterized by a mass function with a systematic radial variation \citep{Andreuzzi2004} expected to depend on the prior dynamical evolution of the cluster \citep{2008AJ....135.2129H}.

Because of its relative proximity, NGC\,6397 has been the focus of several observational and theoretical studies, as well as N-body simulations \citep{2008AJ....135.2129H}. At a distance from the Sun of $R_{ss}= 2.2^{+0.5}_{-0.7}$ kpc \citep{2012ApJ...761...51H}, NGC\,6397 is one of the nearest GCs, which makes its faint stars more accessible than in other more distant GCs. Besides, NGC\,6397 has a very-low metallicity, $[Fe/H] = -1.99\pm 0.02$ \citep{2009A&A...508..695C}, which indicates that it was formed from primordial clouds in the Milky Way, or elsewhere.

Regarding mass segregation in NGC\,6397, previous works present conflicting results, with some showing evidence of mass segregation \citep[e.g.][]{1995ApJ...452L..33K,Andreuzzi2004}, while others do not \citep[e.g.][]{1993MNRAS.265..773D}.

In this work we present a photometric analysis of NGC\,6397 to quantify mass segregation. The paper is structured as follows: in Sect.~\ref{N6397} we present previous and relevant information on NGC\,6397 regarding mass segregation; in Sect.~\ref{SMS} we describe the observational data reduction; in Sect.~\ref{Lumifun}, we present the preliminary analysis of projected data; in Sect.~\ref{Deproj} we build the deprojected profiles; in Sect.~\ref{Discus} we discuss the results and present our concluding remarks.

\section{Previous analysis on NGC\,6397}
\label{N6397}

NGC\,6397 is located at $\alpha = 17^h40^m42.09^s$ and $\delta = -53\degr40\arcmin27.6\arcsec$ (J2000), at a distance from the Galactic centre of $R_{gc}$ = 6\,kpc [\cite{Harris1996} (2010 edition)]. Its galactocentric coordinates are $\ell = 338.17\degr$ and $b = -11.96\degr$, which explains the relatively low contamination by field stars. Its absolute distance modulus is (m-M)$_{0}$ = 12.07$\pm$ 0.06 \citep{2013ApJ...778..104R}, A$_{V}$ = 0.56 \citep{2012ApJ...761...51H} using R$_{V}$ = 0.31, according \cite{1989ApJ...345..245C}. Despite its proximity, the innermost part of the cluster is relatively hard to observe due to the high stellar density. Indeed, NGC\,6397 is classified as a \emph{core collapse} cluster, which has been confirmed by observations \citep{Trager1995,DjorgovskiKing1986}. Using Hubble Space Telescope (HST) data, \cite{1997AJ....114.1517S} estimates an angular size of $\sim5.5$ arcsec for the collapsed core of NGC\,6397, which agrees with \cite{Cohn2010}, but somewhat larger than 3\,arcsec found by \cite{Harris1996}.

Using the observed proper motion of main-sequence stars, \cite{2012ApJ...761...51H} estimated the kinematic mass of the cluster as $(1.1 \pm 0.1) \times 10^{5} M_{\odot} $. They show (fig.~7), through of the mean projected radius of main sequence stars, an apparent mass segregation up to 5\,arcmin. They estimated that NGC\,6397 has about 200\,000\,stars, with $2.5$ per cent of binaries, and visible white dwarfs with masses between 0.5 and 0.6 $M_{\odot}$ \citep{2013ApJ...778..104R}. \cite{1993MNRAS.265..773D} used models of surface brightness profiles (SBP) to find a total mass $7.5 \times 10^{5} M_{\odot} $ and tidal radius of 39.1\,pc.

\cite{1995ApJ...452L..33K}, imaging the highly-concentrated core of NGC\,6397, found that mass segregation effects were very large, compared with the small degree of segregation determined by \cite{1993AJ....106.2335D}. \cite{1995ApJ...452L..33K} observed with the Wide Field Planetary Camera 2 (WFPC2) a field taken 4.5\,arcsec from the cluster centre. Adjusting a King model to the projected data, they fitted a proper dynamical model of the cluster to verify the amount of segregation. 

\cite{1998ApJ...508L..75C} suggested that the strong concentration of cataclysmic variables and a new class of faint UV stars towards the cluster centre is inconsistent with mass segregation related to the two-body relaxation alone seen by \cite{1995ApJ...452L..33K}.

\cite{Andreuzzi2004} built completeness-corrected colour-magnitude diagrams (CMDs) and luminosity functions (LFs) for main-sequence (MS) stars extracted from two fields extending from a region near the centre of the cluster out to $\sim10$\,arcmin. They found that the LFs follow exponential laws with different slopes, being flatter near the centre than otherwise. This is consistent with the presence of different mass distributions at two different radial distances, a clear indication of mass segregation. 

The internal dynamics of NGC\,6397 was addressed by \cite{2012ApJ...761...51H} by means of proper motion measurements obtained with HST's Advanced Camera for Surveys, revealing an agreement with the features of a mass-segregated, lowered isothermal distribution of stars, especially for the selected MS subsamples. 

\cite{2013ApJ...778...57G} quantified mass segregation in 54 Milky Way GCs, including NGC\,6397, by fitting models to cumulative projected star count distributions. The work expands a database of structural and dynamical  properties for 153 spatially resolved star clusters presented by \cite{2005ApJS..161..304M}, where the structural and dynamical parameters were derived from fitting three different models. The results were compiled in \cite{Harris1996} (2010 edition). \cite{2013ApJ...778...57G} calculated the projected density distribution for various boundary conditions and found that fitting star counts rather than surface brightness profiles produce results that differ significantly due to the presence of mass segregation.

\section{Observational data and reduction}
\label{SMS}

The images of NGC\,6397 used in this work are part of the program ID 083.D-0653(A) and were taken in 2009 July 27-28, with the ESO-VLT-UT1 telescope (\textit{Very Large Telescope}), using FORS2 (\textit{FOcal Reducer and low dispersion Spectrograph}) imager. The bands, filters and time exposures are described in Table~\ref{exposure}. We obtained one image for each filter and exposure time. The FORS2 imager is composed by two CCDs (CCD1 and CCD2) of $2\,048\times2\,068$ pixels and scale $0.25"$/pixel. The usable portion of the images contains $1\,670\times1\,677$ pixels.

\begin{table}
    \caption[Observational data table]{ESO-VLT observational data, used in the CMD and the projected Luminosity Functions.}
    \label{exposure}
    \vspace{1em}
    \centering
    \begin{tabular}{c c c}
    Band & Filter &  Exposure time  (s)  \\
    \hline 
    B & B\_HIGH & 1, 10, 120, 300 \\
    V & V\_HIGH &  1, 10, 120, 300 \\
    \hline
    \end {tabular}
\end{table}

\begin{figure}
\centering
\includegraphics[width=84mm,clip=]{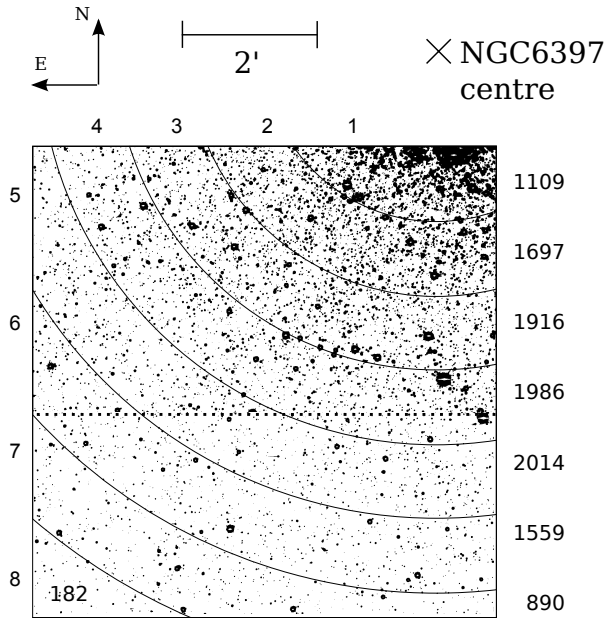}
\caption{Rings around the centre of NGC\,6397, superposed to the ESO-VLT image. The inner ring lies at $\sim1$\,arcmin from the cluster centre (0.72\,pc); the approximate ring width is 0.74\,pc. The number of stars in each ring is shown at the right, except for the last ring (inside), as well as the ring number (above and left). The dotted line indicates the border between CCD1 (above) and CCD2 (below).}
\label{rings}
\end{figure}

CCD1 and CCD2 cover fields of $6.8\times3.9$\,arcmin and $6.8\times2.9$\,arcmin, respectively, amounting to a total field of $6.8\times6.8$\,arcmin, without any gap between them. The images were centred at $\alpha = 17^h41^m01.78^s$ and $\delta = -53\degr44\arcmin47.2\arcsec$ (J2000) as shown in Fig.~\ref{rings}, corresponding to $1.1$\,arcmin from the cluster centre to the upper right corner of CCD1.

Data reduction for our ESO-VLT images were carried out with \textsc{iraf} routines, with tasks \textit{daofind, phot} and \textit{allstar}, applying standard PSF photometric procedures. The best-fitting PSF function among those available in \textsc{iraf} was the elliptical Moffat with coefficient $\beta = 2.5$. The precise description of the PSF is important, for instance, for the insertion procedure of artificial stars (Sect.~\ref{ArtStar}). Objects brighter than V $\simeq $ 19 were saturated in the 300\,s images, thus both data sets were combined to avoid superposition. We detected 12\,793 sources in CCD1 and 6\,528 in CCD2, totalling 19\,321 detections. The photometric uncertainties were computed automatically by \textsc{iraf}.

\section{Preliminary analysis}
\label{Lumifun}

The first step in our study of mass segregation is to build the LF with our ESO-VLT data. We observed the LF difference between the region closest to the cluster centre and the outer region. It is necessary to make completeness analysis, being a crowded cluster, to determine whether this difference is observed only due to completeness effects of the data or it is the effect of real mass segregation.

\subsection{Colour-magnitude Filter}

\begin{figure}
\centering
\includegraphics[width=84mm,clip=]{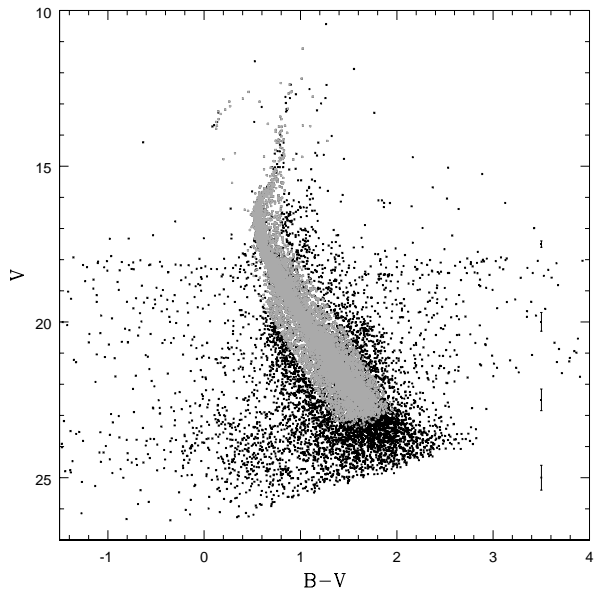}
\caption{CMD of NGC\,6397, with the Tabl.~\ref{exposure} images. The cluster members are shown in gray and the rejected - filtered out - stars in black, for our ESO-VLT data. The bars on the right represent $\pm1\,\sigma$ photometric uncertainties in V magnitude. This diagram was used to make the selection of the cluster stars.}
\label{cmd}
\end{figure}

To exclude obvious non-member sources (mainly field stars), we use a colour-magnitude filter applied to the CMD built with the ESO-VLT data (Fig.~\ref{cmd}), taking the photometric uncertainties into account. Obviously, this procedure also eliminates some stars that belong to the cluster (e.g. white dwarfs), but these are practically negligible for our purposes. This procedure eliminated $\sim$ 25 per cent of the stars originally present in the CMD of NGC\,6397. 

\subsection{Image Slicing}

In order to investigate the spatial dependence of properties such as the luminosity function and mass segregation, the ESO-VLT images were sliced into 8 radial concentric sections with the same width, centred at the cluster centre (Fig.~\ref{rings}). Hereafter, these surfaces are referred to as \emph{rings}.

The rings are numbered from 1 to 8, from near the cluster centre outwards. The first ring is placed at $\sim1$\,arcmin = 0.72\,pc from the cluster centre, with a radial 
width of 0.74\,pc, using the distance to the cluster given by \cite{2013ApJ...778..104R}. The number of probable-member stars in each ring is shown in the right side of Fig.~\ref{rings}.

\subsection{Completeness Correction}
\label{ArtStar}

Before investigating the spatial dependence of the LFs, we first study the photometric and spatial completeness of our data by means of artificial star simulations. In summary, we insert, just in the 300\,s V-filter images, grids of identical artificial stars (equally spaced and with the same magnitude at each time). This V magnitude ranges from 13 to 26, with increments of 0.5 magnitude. The grids of artificial stars for CCD1 and CCD2 contain 1\,008 and 756 stars, respectively, which corresponds to $\cong$ 10 per cent of the total number of observed stars in the images. These numbers are expected high enough to produce statistically significant completeness profiles for each ring without leading to excessive crowding.

\begin{figure}
\centering
\includegraphics[width=84mm,clip=]{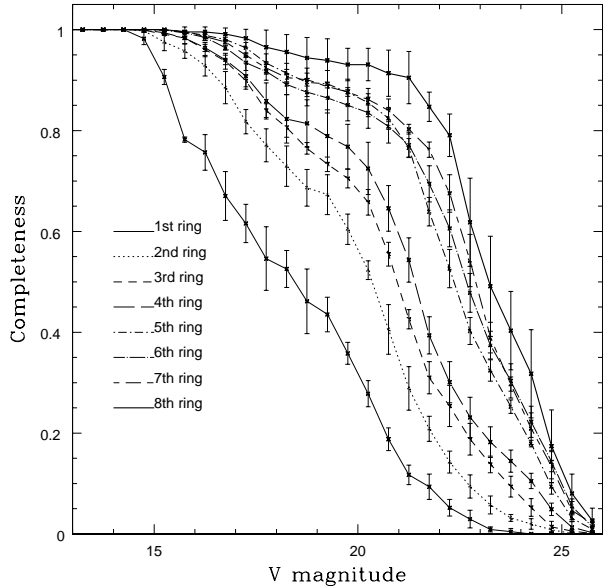}
\caption{Completeness in V calculated for each ring. The bars represent $\pm$ 1\,$\sigma$ uncertainties.}
\label{compincert}
\end{figure}

The artificial stars were created using the same PSF from the data reduction (Sect.~\ref{SMS}) through the task \textit{imexpr} from \textsc{iraf}. Then, they were added to the actual images (task \textit{imarith}) and reduced, to select and measure the position and magnitude of the stars, following the same procedures as for the original images (Sect.~\ref{SMS}). We define completeness as the fraction of stars recovered with maximum deviation of 0.25 magnitudes relative to the inserted value (brightness test), and with maximum deviation of one pixel in position (positional test). Thus, we can estimate the completeness as a function for magnitude for each ring.

\begin{figure}
\centering
\includegraphics[width=84mm,clip=]{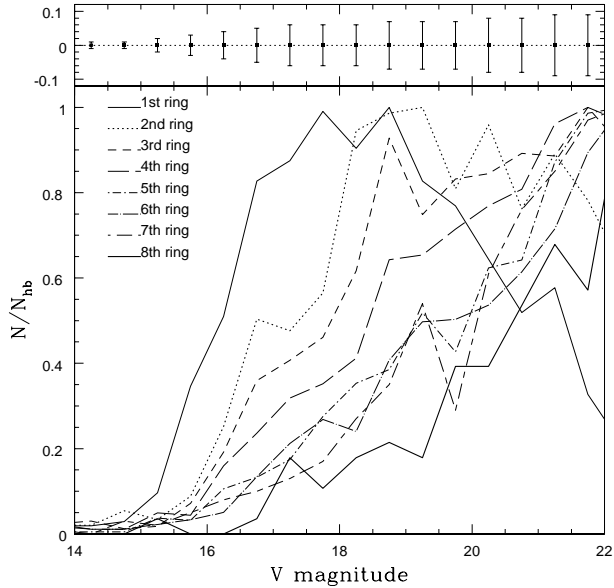}
\caption{Luminosity function for each ring in V, normalized by the highest bin (hb), still not corrected by completeness. Representative uncertainties in $N/N_{hb}$ as function of magnitude are shown in the top panel.}
\label{hist}
\end{figure}

In the simulations the grids of artificial stars were placed at five different positions, shifting the stars in $\Delta x = 5$ and $\Delta y = 5$ pixels from the initial position. After obtaining the completeness for all grid positios, we computed the average completeness and its variance (taken as the uncertainty). The completeness as function of magnitude (and uncertainties) for each ring are shown in Fig.~\ref{compincert}.

\subsection{Luminosity Function} 

The LF is an important information source on several features of a GC, especially those related to the dynamical state and evolution. In particular, a radial dependence of the LF might imply mass segregation \citep{1997MNRAS.286.1012F}.

\begin{figure}
\centering
\includegraphics[width=84mm,clip=]{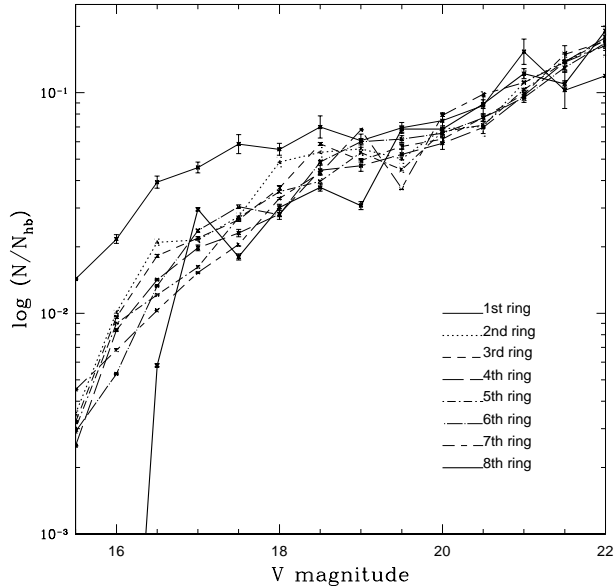}
\caption{Luminosity function for each ring in V normalized by the highest bin with the completeness-corrected LF. Bright stars are more frequent near the cluster centre than outwards.}
\label{nmag8anel}
\end{figure}

We built the LF of NGC\,6397 by dividing the selected stars from its CMD (Fig.~\ref{cmd}) into bins of 0.5\,mag wide in V magnitude, both before (Fig.~\ref{hist}) and after (Fig.~\ref{nmag8anel}) the completeness correction. The LF is expressed in terms of the number of stars per magnitude bin. In this case, normalized by the highest bin. Interestingly, the LFs show systematic differences consistent with mass segregation (especially the increasing presence of bright stars towards the inner rings) before completeness correction.

After completeness correction, the resulting LFs (Fig.~\ref{nmag8anel}) confirm the presence of a higher density of brightness (V $<$ 18), thus more massive stars, towards the central region of the cluster, which is consistent with mass segregation. At the same time, there is a marked deficiency of bright stars in the 
outermost 8th ring. 

However, projection effects may be affecting the observed profiles. For instance, a bright star (V $ < $ 18) detected in the central region may, in fact, be an outsider projected therein, thus masking mass segregation. To investigate this point, we study the mass profile by means of deprojecting the luminosity and density profiles.

\section{Deprojection Profiles}
\label{Deproj}

The rings discussed in the previous section correspond to on-the-sky projections of a three-dimensional structure. In this sense, the projected LFs of a given ring are {\em contaminated} by stars from the outer shells, in fractions that depend on the intrinsic (stellar or brightness) density distribution and the ring width. 

Thus, before performing the final analysis of mass segregation in NGC\,6397, we will proceed to deproject both the surface brightness profile (SBP) and the stellar radial density profile (RDP). Here, we define \emph{R} as the projected distance from the cluster centre (in arcsec) as seen by the viewer, and \emph{r} as the real distance (non-projected) from a three-dimensional coordinate to the cluster centre. 

Since our observations are spatially restricted, not including the central region and outskirts of the cluster, we complement them with data available in the literature, combining CCD and photographic observations. 

\subsection{Surface brightness profile}
\label{SBP}

\begin{figure}
\centering
\includegraphics[width=84mm,clip=]{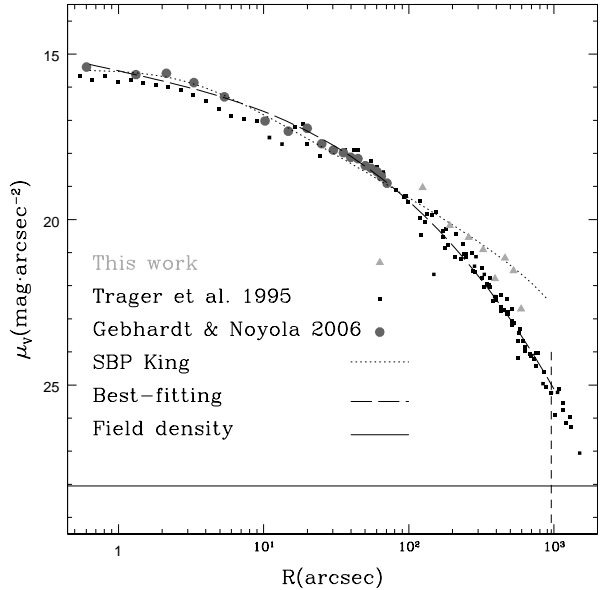}
\caption{Surface brightness profile, the fitted function and the data used in this work to determine the cluster luminosity density. The King profile (dotted line) for the luminosity does not agree with the observed profile. The best-fitting used in this work is the Chebyshev polynomials (dashed line), similar to those used by Trager et al. (1995). The vertical-dashed line indicates the tidal radius and the solid line the luminosity-density of Milky Way field stars.}
\label{trager}
\end{figure}

We compared three SBPs of NGC\,6397: the profile proposed with a King profile \citep{King1962,1966AJ.....71...64K}, the data observed by \cite{Trager1995} and by \cite{2006AJ....132..447N}. However, as it is shown in Fig.~\ref{trager}, the SBP of NGC\,6397 does not follow a King profile, especially for R $>$ 1\,arcmin. We do not use the ESO-VLT data for the SBP because our data correspond to less than a quarter of surface rings, covering from 1\,arcmin out to 10\,arcmin.

The data of \cite{Trager1995} cover the entire cluster in the V filter (Fig.~\ref{trager}), reaching out to 16\,arcmin from the cluster centre, while \cite{2006AJ....132..447N} cover the centre to $\sim$100\,arcsec in converted V magnitude (using WFPC2/HST images in F555W, F606W and F814W). The latter SBP diverges somewhat from that of \cite{Trager1995} only in the cluster centre (probably improvement due to CCD images). These more recent data (out to 100\,arcsec) and \cite{Trager1995}'s data (from 100\,arcsec out to the tidal radius) are used to fit the SBP in this work using Chebyshev polynomials according to

\begin{equation}
\label{mu1}
\mu_{V} = 19.022\,T_0(x) + 4.894\,T_1(x)+ \nonumber \\
\end{equation}
\begin{equation}
\label{mu2}
1.402\,T_2(x) + 0.293\,T_3(x)   
\end{equation}
where $x=\log$(R/arcsec) and $T_n(x)$ is the \emph{n}-degree Chebyshev polynomial, evaluated within the range $[-1,1]$. The behaviour of the proposed functions is shown in Fig.~\ref{trager}. Chebyshev polynomials are used together with the King profile fitted to the RDP to determine the average luminosity radial function.

We use TRILEGAL \citep{Girardi2005} to estimate the luminosity and density of Milky Way field stars in 
the direction of NGC\,6397, obtaining a V surface brightness of 28.05 (arcsec$^{-2}$), and a contamination by Galactic stars of 0.04 stars $\cdot$ arcsec$^{-2}$. These values are lower than those reached at the outermost RDP (Fig.~\ref{dacosta}) and SBP (Fig.~\ref{trager}) bins. The extended function SPB \citep{Trager1995} reaches the Milky Way luminosity density at 950\,arcsec, the value we adopted for the tidal radius.

\subsection{Radial density profile}
\label{RDP}

\begin{figure}
\centering
\includegraphics[width=84mm,clip=]{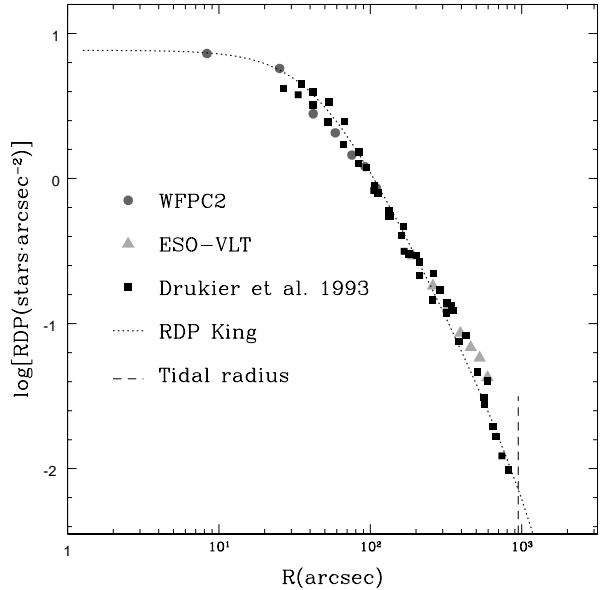}
\caption{Radial density profile (RDP) of NGC\,6397, using Drukier et al.(1993) data (squares), shifted ESO-VLT data (triangles) and WFPC2 data (dots), fitted to a King profile (dotted line). The vertical-dashed line indicates the tidal radius.}
\label{dacosta}
\end{figure}

The RDP (Fig.~\ref{dacosta}) corresponds to the merging of three different data sets: \cite{1993AJ....106.2335D} data covering the whole extent of the cluster, the WFPC2/HST images of the public file of the Hubble Legacy Archive that cover the central region ($R\le60$\,arcsec) and our ESO-VLT (Sect.~\ref{SMS}) for the outer parts ($60$\,arcsec $\le R \le 600$\,arcsec).

First, we matched the outermost point (950\,arcsec) in the \cite{1993AJ....106.2335D} RDP to the background density predicted by TRILEGAL, assuming that missing stars (fainter than the observational data) form a constant distribution in the field, multiplying the image stars by a constant factor. Then, we matched the profiles smoothly under the constraint that the total amount of stars corresponds to a mass equal to $1.1 \times 10^{5} M_{\odot}$ \citep{2012ApJ...761...51H}. 

We find that the total number of stars in the cluster is about 193\,000$\pm$19\,000 down to the limit of hydrogen burning. \cite{2012ApJ...761...51H} suggest a cluster with more than 150\,000 stars. \cite{2008AJ....135.2129H} performed a model N-body of star cluster evolution starting with 100\,000 stars (due to computational limitation) and suggest to perform with twice the number of stars to be more realistic. 

The resulting RDP follows quite closely the King profile~\citep{King1962}:

\begin{equation}
\label{RDP}
RDP(R) = \Sigma_{BG} + \Sigma_0 {\Bigg\lbrace\frac{1}{[1+(\frac{R}{R_{CD}})^2]}-\frac{1}{[1+(\frac{R_{TD}}{R_{CD}})^2]}}\Bigg\rbrace^{\frac{1}{2}}
\end{equation}

By fitting this function to the model by \cite{1993AJ....106.2335D} to the ESO-VLT and WFPC2 RDP (Fig.~\ref{dacosta}) and using as constraint the total mass [$(1.1 \pm 0.1) \times 10^{5} M_{\odot}$ \citep{2012ApJ...761...51H}] as well as the SBP, we obtained the values $\Sigma_{BG} = 0.04$ (stars$\cdot$arcsec$^{-2}$) for the field-star density, $\Sigma_0 = 6.78$ (stars$\cdot$arcsec$^{-2}$) for the density in the cluster centre, $R_{CD} = 42.0$\,arcsec for the core radius-density and $R_{TD} = 3\,000$\,arcsec and for the tidal radius-density. 

We use the units for the deprojected luminosity density and the density-number (the ratio between the magnitude and the number of stars by a volume) equal to one V magnitude$\cdot$arcsec$^{-3}$ and stars$\cdot$arcsec$^{-3}$, respectively.

\begin{figure*}
\begin{minipage}{115mm}
\centering
\includegraphics[width=115mm,clip=]{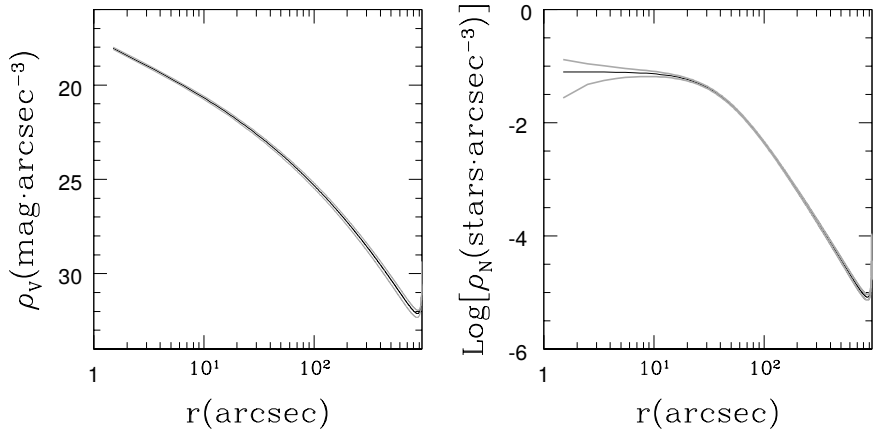}
\caption{Deprojected luminosity density $\rho_V(r)$ in V magnitude (left panel) and density number $\rho_N(r)$ (right panel) as a radial function of $r$. These curves result from the geometric deprojection of the radial ($R$) data (Fig.~\ref{M_r1}).}
\label{M_r2}
\end{minipage}
\end{figure*}

\begin{figure*}
\begin{minipage}{115mm}
\centering
\includegraphics[width=115mm,clip=]{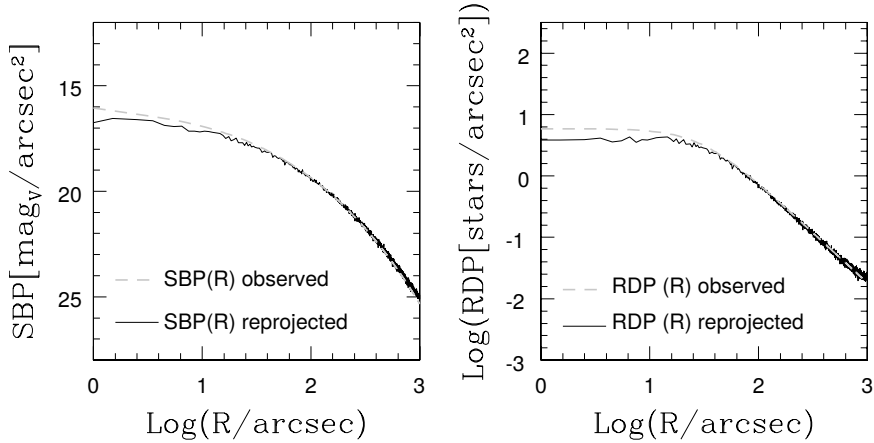}
\caption{Surface brightness profile in V magnitude (left panel) and radial density number (right panel) as a radial function $R$. The dashed lines represent the input functions (SBP and RDP) and the solid lines represent the deprojected and reprojected profiles. The data fit from the tidal radius down to 1\,arcsec for both functions.}
\label{M_r1}
\end{minipage}
\end{figure*}

To determine the deprojected number-density of cluster stars [$\rho_N(r)$ in Fig.~\ref{M_r2}], we calculate the volume of the outermost ring (geometrically) and its number-density (from observational data), computing this amount for the outermost shell. Thereafter, we calculate the number of stars within the immediate inner ring and its volume. Subtracting the quantity in density of the outer rings, we estimate the density of the ring by dividing the remainder number of stars by its calculated volume. Thus, the density for each shell is computed from the values ​​of the densities of the outer shells. The same procedure is used for the brightness density [$\rho_V(r)$ in Fig.~\ref{M_r2}]. We use a 1\,arcsec step for the ring width. We employed geometrical deprojections instead of an inverse Abel transform because the former method can be used with discrete data (stars). Geometrical deprojection was suitable for our purposes and reprojection agrees with observational data (Fig.~\ref{M_r1}). 

Subsequently, we projected the cluster stars assuming a radial symmetry in their distribution. In order to test it, we compared the reprojected data to the observed profiles in Fig.~\ref{M_r1}, showing the original projected profiles SBP (from Fig.~\ref{trager}) and RDP (Sec.\ref{RDP}) in dashed blue lines. Both functions fit quite well the reprojected data (solid black lines) from the tidal radius down to the cluster centre. 

Calculating the ratio between the radial deprojected luminosity density $\rho_V(r)$ (Fig.~\ref{M_r2}, left panel) by the radial deprojected density-number $\rho_N(r)$ (Fig.~\ref{M_r2}, right panel), we established the averaged luminosity radial function (magnitude $\bar{V}$ in Fig.~\ref{M_r3}) of the stars. 

To convert luminosity (Fig.~\ref{M_r3}) to mass (Fig.~\ref{M_r4}), we used the best fitting multichromatic isochrone. The parameters were obtained by a statistical method comparing the mean ridge line from CMDs using multiple colours from the grid of isochrones from the Dartmouth Stellar Evolution Database~\citep{2008ApJS..178...89D}. The best-fitting parameters for NGC 6397 are an age of 12.0$\pm$0.5 Gyr, [Fe/H] = - 1.8 $\pm$ 0.1, E(B-V) = 0.12 $\pm$ 0.01 magnitudes and distance modulus (m-M)$_{0}$ = 12.04 $\pm$ 0.01 magnitudes.

\section{Summary and discussions}
\label{Discus}

The completeness-corrected projected luminosity function built with our ESO-VLT photometry suggests the presence of mass segregation in NGC\,6397. Indeed Fig.~\ref{nmag8anel} clearly shows a higher density of brighter stars near the projected central region of the cluster with respect to the outer parts. This apparent evidence may be due to contamination by outer shell stars.

Because of projection effects, mass segregation may not be directly detected on-the-sky luminosity functions. In this sense, geometrical deprojection appears to be a better approach. The analysis of a non-segregated cluster may show a crowded projected central field where the fainter stars are not detected, unlike what happens near the tidal radius. 

To confirm the presence (or not) of mass segregation, we deprojected both the surface brightness and radial density profiles of NGC\,6397. The SBP was built with data by \cite{Trager1995} and \cite{2006AJ....132..447N} (Fig.~\ref{trager}). For the radial density profile, we used a merging of three different data sets: ~\cite{1993AJ....106.2335D}, WFPC2/HST and ESO-VLT data (Fig.~\ref{dacosta}). We obtained then, the deprojected SBP and RDP, using their respective profiles (Fig.~\ref{M_r1}) and deprojecting both these functions. 

\begin{figure}
\centering
\includegraphics[width=84mm,clip=]{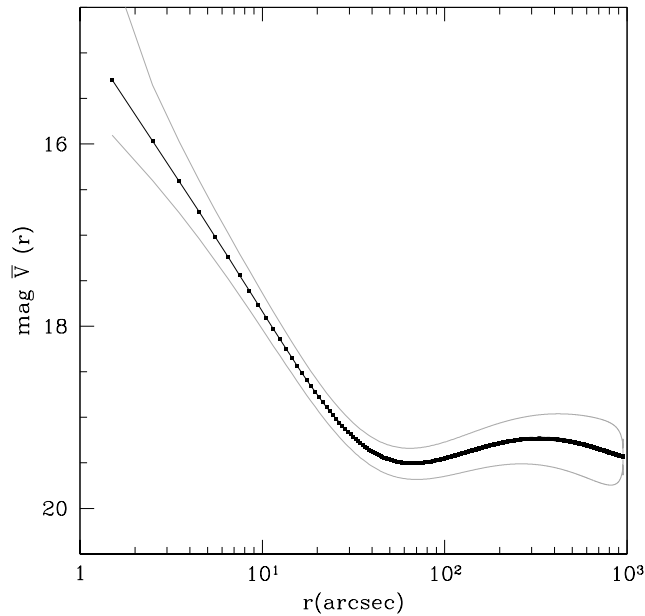}
\caption{The ratio between $\rho_V(r)$ and $\rho_N(r)$ leads to the averaged magnitude $\bar V$ of the cluster stars, following a decreasing function from about 30\,arcsec out to the tidal radius. The gray lines show the $1\pm\sigma$ uncertainties calculated using both brightness and star count errors.} 
\label{M_r3}
\end{figure}

Finally, computing the ratio between the deprojected radial luminosity density by the deprojected radial number-density (Fig.~\ref{M_r2}), we obtained the averaged radial luminosity function (Fig.~\ref{M_r3}). We also determined the cluster radial mass function.

\begin{figure}
\centering
\includegraphics[width=84mm,clip=]{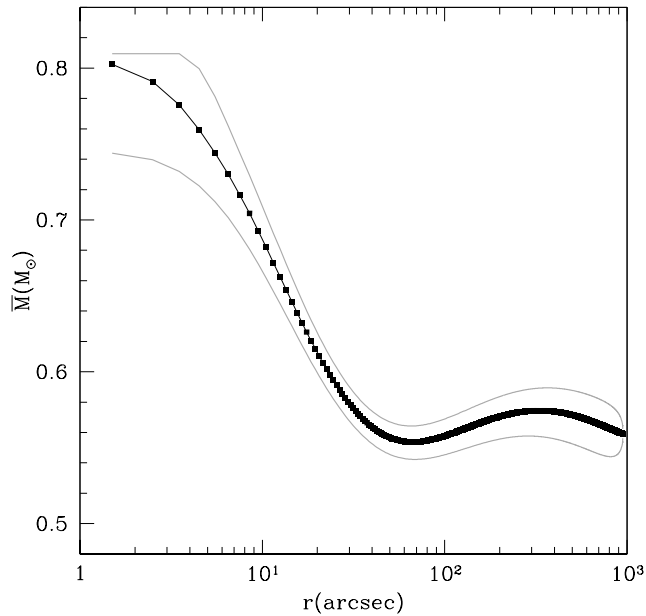}
\caption{The luminosity function was converted to mass (in units of solar masses) using the best fitting isochrones \citep{Pieres2014}. The gray lines show the  $1\pm\sigma$ uncertainties calculated using both brightness and star count errors. The upper limit to the mass uncertainty are imposed by mass implied by the isochrone.} 
\label{M_r4}
\end{figure}

Figure~\ref{M_r4} shows that the mass segregation occurs basically in the centre of the cluster reaching up to a radius of 1\,arcmin. In other words, we confirm the presence of mass segregation using deprojected data. Other authors~\citep[e.g.][]{1995ApJ...452L..33K,1998ApJ...508L..75C,Andreuzzi2004,2012ApJ...761...51H} evinced mass segregation to different radius using just projected data.

Even though the projected data in Figs.~\ref{hist} and \ref{nmag8anel} and the literature [e.g.~fig~7 \cite{2012ApJ...761...51H}] show evidence of mass segregation beyond 1\,arcmin, the deprojected profiles (Figs.~\ref{M_r3} and \ref{M_r4}) do not. The reason is that deprojection considers the contribution of the outer regions to the inner ones, thus, we compute the individual uncertainty for each ring by means of error propagation, as are shown in Figs.~\ref{M_r2}, \ref{M_r3} and \ref{M_r4}. The segregation in the centre of mass is well established. So, the projected functions (RDP and SBP) are still accurate, showing that there is no mass segregation evidence in the deprojected profiles beyond 1\,arcmin.

We also observed a soft depression around 0.55\,M$_{\odot}$ (radius equal to 1\,arcmin) in Fig.~\ref{M_r4}. The decrease in this region can be explained by molecular absorption in the stellar atmospheres, also observed in the luminosity function \citep{Richer2008}. They explain that the mass-brightness relation is modified by molecular absorption in such a way that a moderate range in stellar mass creates a small range in luminosity so that the stars accumulate at this luminosity.

The conversion of luminosity into mass shows a degenerescence between white dwarf masses [$\bar M \cong $ 0.55\,$M_{\odot}$] and main sequence star masses [$\bar M \cong $ 0.15\,$M_{\odot}$] for the same luminosity (V $\cong$ 24). The white dwarf stars were neglected in this approach, being the main error source for the determination of mass, although the number of white dwarfs should be smaller than in main sequence stars.

In the present study, we found a way to determine the total number of stars in the cluster using the estimated mass as an input parameter. These are important results for theoretical studies, as well as N-body simulations.

The presence of dust is negligible and the luminous mass of the cluster is enough to explain the gravitational energy of the most distant stars, although mass percentage (uncertain) could be due to non-luminous mass. On the other hand, assumptions about dark matter (large amounts as in the galaxies) in the case of NGC\,6397 are irrelevant \citep[e.g.][]{2013JKAS...46..173S}.

\section*{Acknowledgments}

We thank the referee for important comments and suggestions.
We thank the ESO-VLT support team for help with the data acquisition.
We acknowledge financial support from the Brazilian Institution CNPq and CAPES.

\end{document}